# Are consumers ready to pay extra for crowd-shipping e-groceries and why? A hybrid choice analysis for developing economies

Oleksandr Rossolov[1], Yusak O. Susilo[1]

[1]Institut für Verkehrswesen (IVe), University of Natural Resources and Life Sciences,
Peter-Jordan-Straße, 82, 1190 Wien, Austria
email: oleksandr.rossolov@boku.ac.at; yusak.susilo@boku.ac.at
O. Rossolov ORCID: 0000-0003-1495-0173, Y.O. Susilo: 0000-0001-7124-7164

**Abstract**
This paper presents the behavioral study's results on willingness-to-pay the extra money by the customers for e-groceries deliveries based on crowd-shipping. The proposed methodology was tested for Ukraine, i.e., a developing country where the crowd-shipping services are under development conditions. To account for the behavior complexity of the consumers who have not faced the crowd-shipping services in the past, the choice model was enhanced with a latent variable. The findings indicate the revealed readiness of the e-shoppers to pay extra money for crowd-shipping delivery if it provides more flexible and consumer-oriented service. The expected environmental impact of the crowd-shipping delivery was not considered as important by the e-shoppers, which is explained by low concerns about the environment and car-oriented mobility in the considered case study.

## 1. Introduction

With the rise of connectivity and e-commerce, the crowd-shipping (CS) practice has attracted a lot of interest in the last few years (for examples of recent works can be seen at Rougès and Montreuil, 2014; Wang et al., 2016; Frehe et al., 2017; Kafle et al., 2017; Marcucci et al., 2017; Punel and Stathopoulos, 2017; Ermagun and Stathopoulos, 2018; Arslan et al., 2019; Gatta et al., 2019; Le and Ukkusuri, 2019; van Duin et al., 2019; Buldeo Rai et al., 2021b; Neudoerfer et al., 2021; Le et al., 2021; Wicaksono et al., 2021; Boysen et al., 2022; Fessler et al., 2022; Seghezzi and Mangiaracina, 2022; Rechavi and Toch, 2022). It is considered as one of the options to promote more sustainable and environmentally friendly last-mile deliveries. Furthermore, recent restrictions and lockdowns imposed due to COVID-19 has accelerates the acceptance and adoption of new shopping channels like e-groceries, not only in well-connected countries, but also in many other parts of the world (Susilo et al., 2021; Maltese et al., 2021; Bin et al., 2021; Rossolov et al., 2022). Having such sudden increase trend, the last-mile logistics services faced with lack of commercial vehicles which was perceived very sharply during the COVID-19 lockdowns (Andruetto et al., 2022). One of the common solutions to address this problem is by using CS-based delivery services where common people act as occasional couriers and can transport the goods within their routine trip chains.

In the light of the CS deliveries a list of questions arose that can be distinguished into two main branches. The first one covers all studies devoted to the demand-side of the CS system aiming at evaluation of the consumers behavior and their readiness to use a such service (Devari et al., 2017; Punel and Stathopoulos, 2017; Punel et al., 2018; Gatta et al., 2019; Wicaksono et al. (2021). In turn the supply aspects of the CS-based delivery system are considered as the second branch within the CS topic (Gatta et al., 2019; Le and Ukkusuri, 2019; Wicaksono et al. (2021); Fessler et al., 2022). Both of these branches are valuable for the efficient CS-based delivery services as within the interaction of the demand and supply subsystems the aftermaths of the last-mile logistics can be evaluated. Most of the studies for both branches cover the developed economies where the e-commerce already reached wide deployment and people are ready to use this shopping channel for both experience and search goods (Axhausen and Schmid, 2019). But it is yet known whether the findings from developed countries would also applied in developing countries, e.g. Ukraine, especially for the groceries where consumers were more oriented to in-store shopping (Rossolov et al., 2021).

Addressing behavioral aspect of the CS-based delivery services is critically important because this business model has becoming more and more popular around the globe, including in developing countries. But having different shapes in the consumers' behavior in developed and developing economies, it is important to reveal the decision-making determinants in the developing economies. These findings will give the possibility to predict the potential users of new delivery services that the crowd-shipping paradigm offers. To go beyond the current conditions of last-mile logistics the stated-preference method can be applied to propose e-shoppers' possible scenarios for CS-based services. Having supply and demand equilibrium problem, it is proposed to make the behavioral estimates with overpriced delivery cost. In such a way the potential demand for CS-based services will be revealed with more stable conditions as it is known that reduction of the service cost attracts more clients.

The paper is structured as follows. Section 2 contains the review of the current state-of-the-art for CS deliveries problem. Within Section 3, we develop the study's methodology by formalizing the trade-off conditions, hybrid choice model description, and stated-preference experiment design. Section 4 contains the results of the study, following from the sample description to descriptive statistics, exploratory factor, and discrete choice analyses. Section 5 discusses the study's main findings and emphasizes its limitations. Section 6 contains a summary of the study and future steps to be taken.

## 2. Literature Review

Given the extensive list of studies focused on the CS problem, the literature review will be made by distinguishing studies into demand-side and supply-side oriented within the problem. Considering separately demand-side studies provides the possibility to evaluate the elaborations made in the field of the consumers' behavior regarding CS service and define the gaps with the steps needed. On the other hand, the focus on supply-side of CS is important in terms of the considered modes and delivery attributes evaluated to promote realization of sustainable last-mile deliveries. The comparison analysis results are summarized in Table 1.

Despite a significant number of studies made on CS service only several of them are focused on the revealing the choice decision of e-shoppers to use this delivery channel as consumers (demand-side) and providers (supply-side). Regarding the attributes that were considered to influence the consumers' choice of SC most of the studies focused on socio-demographic data (Devari et al., 2017; Punel and Stathopoulos, 2017; Punel et al., 2018; Gatta et al., 2019; Le and Ukkusuri, 2019; Fessler et al., 2022) which is common for the behavioral research. Among all socio-demographic attributes, the personal or household income is worth noticing as it is considered as the basic parameter to reveal CS-oriented consumers. Thus, Devari et al. (2017) stated that income attribute influence positively on SC choice by e-shoppers meaning the growth of this value will increase the likelihood to use crowd-based deliveries. Contributing to that finding, Punel and Stathopoulos (2017) specified that low-income people less intent to use CS instead of high-income people. Contradicting to that, Punel et al. (2018) estimated for the USA that people with personal annual income greater than 59.000 USD do not prefer CS deliveries. Having such a contradicting relation between consumer's income and his/her willingness to use the CS deliveries, this question is worthy of additional attention especially for developing economies.

Among other socio-demographic attributes, people with low education and employment status "full and part-time" have been evaluated with negative attitudes to CS (Punel ans Stathopoulos, 2017). In contrast to this, full-time consumers reveal pro-CS behavior and are accustomed to that service in the USA (Punel and Stathopoulos, 2017; Punel et al., 2018). Considering gender-related question, the CS services attract more male consumers than females (Punel et al., 2018). From the prospect of consumers' age, Gatta et al. (2019) have observed that senior people are less likely to use CS instead of young e-shoppers.

The impact of the delivery channel attributes on consumer behavior has also been considered within CS-related studies. Thus, Gatta et al. (2019) and Wicaksono et al. (2021) have evaluated in their models the impact of the delivery/shipping rates and delivery/shipping times on delivery channel choice behavior. In CS-related studies, special attention is given to the quality of the delivery services and how it impacts the trade-off conditions. For instance, Devari et al. (2017) revealed that consumers are susceptible to the detour time needed for crowd-shipper for delivery implementation. An increase in detour time reduces the attractiveness of CS services. Complementing this finding, Punel and Stathopoulos (2017), Gatta et al. (2019) evaluated the importance of CS delivery to be flexible, providing e-shoppers with the possibility to trace their parcels and adjust the pick-up times and locations. Along with that, Punel et al. (2018) revealed that people in developed economies expect CS services to be eco-friendly with a positive contribution to society. Such concerns describe a new level of trade-off conditions when people evaluate not only the direct cost to be paid out of pocket but also the external cost.

It should be noted that the above-mentioned studies did not consider the willingness-to-use CS for e-groceries. Some of them focused on parcel deliveries in general (Devari et al., 2017, Gatta et al., 2019) or analyzed CS deliveries for generalized "goods" (Punel and Stathopoulos, 2017; Punel et al., 2018; Wicaksono et al., 2021). Despite that, the obtained findings are a good base for enhancing the choice situation with more detailed quality-related attributes. Applying them to e-groceries' conditions with the CS option provides a new possibility to reveal more real-case conditions on how consumers trade off the cost and quality of service for food supplies.

In this case the considered alternatives in the choice set play an important role in assessing the e-shoppers behavior regarding CS services. Punel and Stathopoulos (2017) using stated-preference method developed a choice set with differentiation of CS-based alternatives in the light of driver features, i.e., occasional drivers versus professional one. Gatta et al. (2019) considered two alternatives as "CS-based delivery" and "No choice" which allowed them to reveal attributes of willingness to use a new delivery service without actual consideration of the trade-off conditions between alternative channels. Devari et al. (2017) modeled similar situation to estimate a binary choice model using logistic regression technique while Punel et al. (2018) used a real case data for the implemented CS deliveries to evaluate determinants of the CS choices made by e-shoppers. Wicaksono et al. (2021) used the stated-preference method and depicted a wider range of alternatives than other studies. The bicycle CS, traditional shipping, and self-pick-up at the physical store are considered options by Wicaksono et al., which reflect a close real-case situation when e-shoppers can choose delivery channels that are presented in the market. From the supply point of view on CS-based deliveries, big attention is given to the socio-demographic features of people that might act as occasional couriers. Le and Ukkusuri (2019) enhanced their model with race specification, which is very important for multicultural and multinational communities. Fessler et al. (2022) evaluated the couriers' age and employment status regarding readiness to provide the CS service, revealing pro-CS-oriented behavior for young adults and employed people. Attention to delivery-related attributes like parcel weight, its size, delivery distance and cost has been given by Ermagun et al. (2020), Wicaksono et al. (2021), and Fessler et al. (2022). Among mentioned studies, Wicaksono et al. (2021) have focused by design on bicycle-based CS deliveries, Gatta et al. (2019) and Fessler et al. (2022) examined public transport, while Ermagun et al. (2020) evaluated private cars usage. Having that, we can summarize that the mode choice problem for CS deliveries has not been studied yet, which is the gap to be covered. Moreover, within a supply-related problem, the question of possible detours that occasional couriers are ready to make within their CS delivery is not yet addressed.

The common feature of the demand-side and supply-side studies can be depicted by the methodologies used. Thus, most of the studies, except Devari et al. (2017) and Punel et al. (2018), are based on the Random Utility Maximization Theory (RUM), which is very efficient, especially for a multi-alternative choice situation. Given that, we plan to develop the behavioral models within this study based on the RUM considering the revealed gaps in this chapter.

Table 1. Basic findings of the behavioral studies in the field of CS service

| Study | Considered scenario(s) | Mode | Studied object | Model type | Revealed attributes of crowd-shipping service choice | Case study/ sample size |
|---|---|---|---|---|---|---|
| Devari et al., 2017 | Friend of e-shopper is supposed to be a receiver | NA | Demand | Logistic Regression | (+) Income<br>(–) Extra time for delivery (detour time) | USA/104 |
| Punel and Stathopoulos, 2017 | Crowd Delivery with different driver options, i.e., professional or occasional driver | Car | Demand | MNL and Mixed MNL | (+) High income; Employed – full-time; schedule pick-up; driver rank<br>(–) Low income; Medium income; Low education; Employed – full and part-time | USA/531 |
| Punel et al., 2018 | Assessment of "crowd-shipping users" vs "non-users" | NA | Demand | Logistic Regression | (+) Eco-Friendly service; Male; Full-time employment; Population density; Contribute to community<br>(–) Help people; Income greater than 59 000 USD | USA/800 |
| Gatta et al., 2019 | Crowd-shipping options based on APL vs "no choice" | Public transport | Demand/ Supply | MNL | (+) Lower shipping fee; Lower shipping time; Presence of parcel tracking; Flexible delivery and schedule time/APL in metro; Remuneration; Real-time booking<br>(–) Age/Age i.e., senior people less intend to use and provide CS service | Italy/206 |
| Le and Ukkusuri, 2019 | Use CS vs Not to Use CS | NA | Supply | Binary Logit | (+) Graduate degree holders; Age; Male; African American/American Indian people<br>(–) Income; Numbers of people in your household | USA/549 |
| Ermagun et al., 2020 | "No Bid" vs "Bid" alternative nests for urban and suburban areas | Private cars | Supply | Nested Logit | (+) Larger Shipment; Job access density<br>(–) Bid Deadline; Holiday; Weekend; Delivery Distance | USA/16,850 |
| Wicaksono et al., 2021 | Bicycle Crowd-shipping vs Traditional Shipping vs Self Pick-up | Bicycle | Demand/ Supply | MNL | (+) Delivery Time Window; Driver Rate; $CO_2$ Reduction / Monetary Compensation<br>(–) Delivery Cost; Delivery Time (Speed) / Package Weight; Additional Travel Time | Netherlands/319 (demand) and 136 (supply) |
| Fessler et al., 2022 | Crowd-shipping vs Not Ready to Provide | Public transport | Supply | Mixed MNL | (+) Age 18-39 year; Employed/Self-employed; Student<br>(–) Age above 60 years; Compensation; Extra Time; Number of Parcels; Size; Weight | Denmark/524 |

## 3. Methodology

To reveal the trade-off between cost and quality aspects of the alternative delivery channels, the higher rates/remuneration to be paid to occasional couriers than to common commercial carriers are considered. Additionally, CS delivery is supposed to provide the service with higher quality than commercial carriers, presented by the instant delivery features with additional flexible options for the consumer to adjust the delivery process. Hence, non-trivial trade-off conditions are considered, reflected by the following question: "*Are people ready to pay higher remuneration for the CS-based delivery service but with a more quick, flexible, and respond-oriented system than the commercial carriers provide?*" As the potential demand for CS service is studied, the decision maker does not have an experience with CS and may not trade off the attributes rationally given pre-concerns against non-depicted-in-experiment attributes. Given that, the leverage of the hybrid choice approach will be used to enhance the choice model with latent variables and describe consumers' choice behavior in a more precise and realistic way (Ben-Akiva et al., 2002).

*3.1 Trade-off conditions*

The CS adoption issue (demand side) is considered within the study from the point of the perceived utility by the consumers due to the choice of specific supply option. The alternative situations of the choices considered within this study are depicted in Fig. 1.

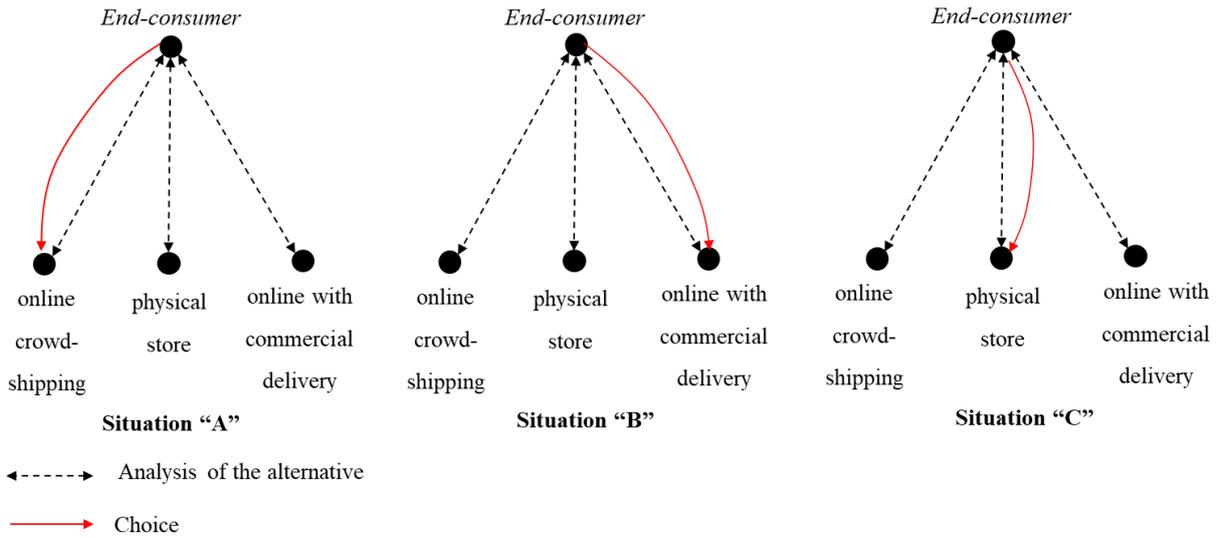

Figure 1. Considered choice situations for groceries supplies

According to the random utility maximization theory (Ben-Akiva and Lerman, 1985; McFadden, 1986) the decision maker (consumer) chooses the alternative *j* if it provides the maximum utility, which mathematically is presented as (Ortuzar and Willumsen, 2011):

$$U_j = \max \forall U_i, i \in A \qquad (1)$$

$$U_i = V_i + \varepsilon_i = \boldsymbol{\beta} \cdot \boldsymbol{X} + \varepsilon_i \qquad (2)$$

where $U_i$ is the perceived utility for alternative *i*, *A* is the number of the alternatives, $V_i$ is the deterministic part of the utility, $\varepsilon_i$ is the error term which is assumed to be i.i.d., $\boldsymbol{\beta}$ is the vector of the parameters (marginal estimates), $X$ is the vector of the attributes.

*Eq. (1)* guarantees that consumer can evaluate/perceive all utilities for all alternatives in the choice set. In turn, *Eq. (2)* formalizes the utilities for every considered alternative. Given that, tree alternatives for groceries supply to be available for end-consumers are distinguished. The first one is the online purchases option with commercial carrier (CC) deliveries which currently presented in the market. The second one is being deployed new service based on CS deliveries for e-groceries. The third supply option reflects the classical way of the groceries shopping – in the physical store which can be appropriate for the people that still are not ready for e-grocery services.

As the problem of online versus in-store grocery shopping has already been studied in last few years (Bjørgen et al., 2021; Gatta et al., 2021; Maltese et al., 2021; Marcucci et al., 2021) this paper focuses on the e-groceries delivery channel choice problem that has been only partially revealed in the studies done so far. The CC and CS-based deliveries are considered in this study as the alternative channels. Given that, CC and CS supply channels should provide the trade-off conditions in terms of some specific features in situations "A" and "B" (Fig. 1) in order to be chosen by consumer. Hence, the trade-off conditions studied are depicted in Fig. 2. As some people are still not ready to buy groceries online, the "Physical store" option is provided in the choice set (Fig. 1) to account for the possibility of rejection all e-groceries channels.

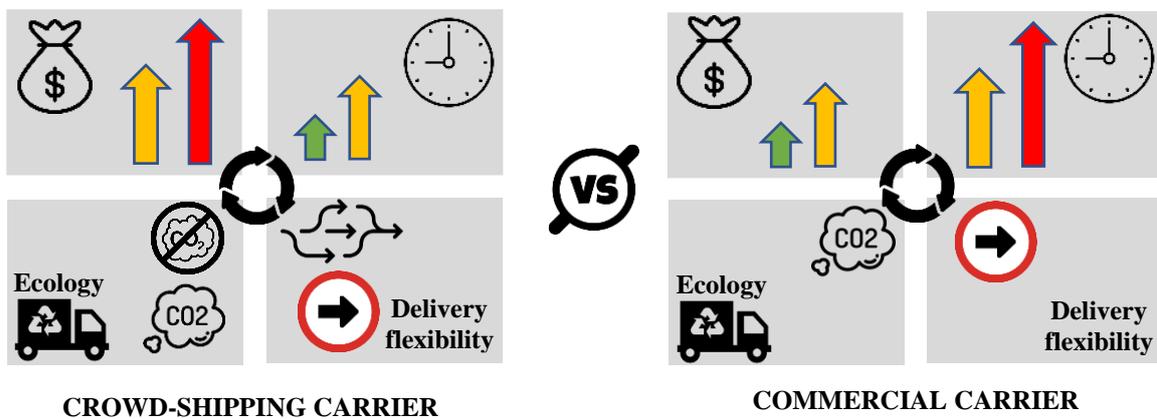

Figure 2. Alternative supply channels' trade-off conditions

According to Fig. 2 we aim to evaluate the following trade-off conditions:
- ***Delivery costs*** are expected to be higher for CS service then for CC allowing us to estimate the readiness to pay more for CS compared to CC services. Given that, the higher costs for delivery are expected to be compensated by the *higher quality* of the service with an optional *ecological* contribution.
- ***Service quality.*** The higher quality of SC-based e-grocery supplies should be guaranteed by the *lower delivery time* and *flexibility* of the system:
  - ***lower delivery time*** for CS service is guaranteed by the low time for the request handling after the bit winning. The same-day delivery for CS-based e-groceries becomes feasible in such conditions. Also, the lowest number of clients within one delivery trip made by SC driver allows them to reduce the delivery time compared with the commercial carrier, which should use a long tour-based delivery to optimize the costs and empty trips (Holguín-Veras and Thorson, 2003; Gonzalez-Calderon et al., 2021);
  - ***flexibility*** supposes that e-shopper is provided with the information exchange channel allowing them to change the delivery point. In this case, the occasional courier can adjust the delivery path along with the time schedule changing as well. In turn, the commercial carrier does not have the possibility to change the delivery address as the delivery route/tour is predefined and the e-shopper is not provided by the communication channel with the commercial driver. This attribute has been included in the choice set following Punel and Stathopoulos (2017) and Gatta et al. (2019) findings.
- ***Ecological*** contribution to be made by CS-based deliveries relies in the following:

- possibility to leverage "green" modes of the private (bike) or public transport (subway tram, and trolleybus), which are more sustainable compared with commercial carriers that mostly employ vans with a combustion engine;
- zero contribution to the traffic by CS deliveries due to the public transport modes usage.

It should be pointed out that the ecological contribution of CS deliveries can be harmful if the courier uses a private car with a combustion engine. In such a situation, we can expect extra pressure on the city environment due to the production of additional traffic, and detour necessity is expected caused by other traffic, emissions, etc. To emphasize that, the "crowd-shipping" alternative in Fig. 2 presents two options for the ecology features, meaning the possible positive and harmful impacts of the CS deliveries. The mechanism of how it is presented in the choice sets will be described in the next chapter, where the methodology for the stated-preference experiment is presented.

*3.2 Hybrid choice model*

Given the considered deploying conditions for CS-based e-groceries, the consumers had yet to face such a service. In fact, the latent demand is studied by exploring *what-if* conditions for the delivery channel choice. Discrete choice models have shown a high efficiency in the marketing and transportation fields when a new product or service is introduced to consumers. Recent studies show that enhancing these models with latent variables allows researchers to reach more detailed estimations (Atasoy et al., 2013). Besides that, incorporating the latent variables into the discrete choice models enables the capture of behavioral complexity, which is more relevant to the choice situations with new products (Guo et al., 2022). Complementing the choice model with latent variable(s) results in the hybrid choice model, which depicted for this research case in Fig. 3.

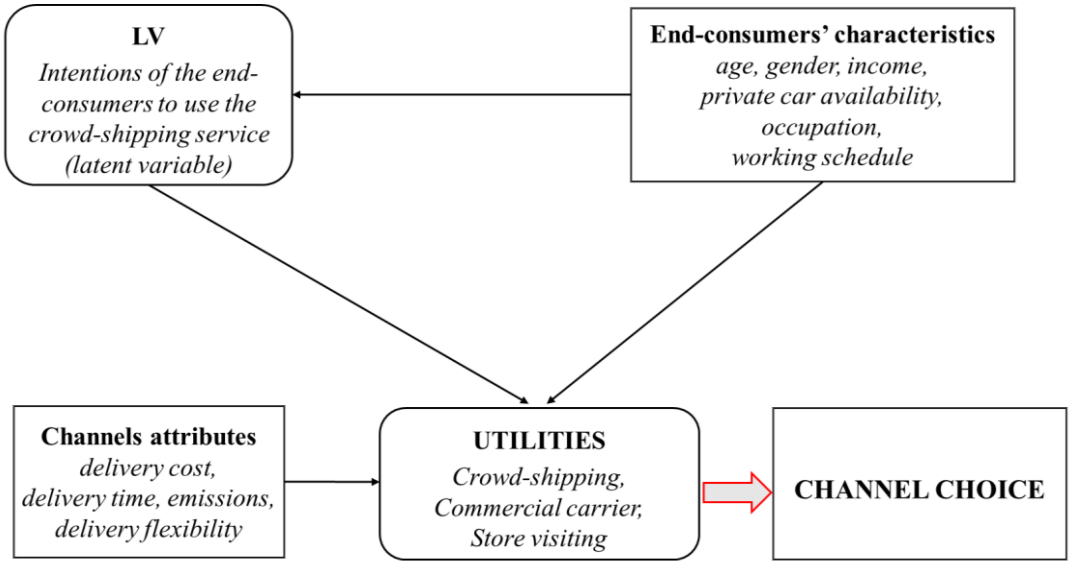

Figure 3. The hybrid choice framework

The first step for the hybrid choice model development supposes the structural equation formalization (Bolduc and Álvarez-Daziano, 2009; Atasoy et al., 2011) as follows:

$$LV_{qz} = h\left(\mathbf{x}_q \mid \boldsymbol{\beta}_q^s\right) + \psi_{LV_{qz}} = \beta_0^s + \sum_{k=1}^{K_s} \beta_k^s \cdot x_k + \psi_{LV_{qz}} = \beta_0^s + \sum_{k=1}^{K_s} \beta_k^s \cdot x_k + \sigma_s \cdot \varepsilon^s \qquad (3)$$

$$\psi_{LV_{qz}} \sim N\left(0, \sigma^2\right) \qquad (4)$$

where $LV_{qz}$ is the structural equation for the latent variable $z$ for consumer $q$, $h(x_q|\beta_q^s)$ is the deterministic part of the structural equation, $\psi_{LV_{qz}}$ is the stochastic part of the structural equation to be normally distributed with zero mean and variance $\sigma^2$, $\beta_q^s$ is the vector of model's parameters to be determined according to vector of the attributes $x_q$ for every consumer $q$, $\beta_0^s$ is the generic constant term, $\beta_k^s$ is the specific model's parameter for $h(x_q|\beta_q^s)$ considering attribute $k$, $x_k$ is the model's attribute.

The technique of indirect estimations is used to reveal the latent variable, which can be done based on the direct or indirect questions with evaluation of them on some scale (Bierlaire, 2018). The five-point Likert scale (Likert, 1932) is the most used, allowing people to shape their attitudes toward estimated statements. This study considers the leverage of the indirect questions technique with a five-point Likert scale to reveal different shapes of the consumers' attitudes towards ecology concerns, trust, social awareness, etc. The factor analysis procedure is the final step for evaluating the latent variable within such a technique. In this case, the structural equation should be formed in accordance with the revealed factors expressing latent attitude. The interconnection between the structural equation and Likert scale estimates is provided by the measurement equations as follows:

$$I_{qw} = m(LV_{qz}, x_q; \beta_w^m) + \upsilon_{qw} = m(LV_{qz}, x_q; \beta_w^m) + \sigma_w^* \cdot \varepsilon_w^*, \; w \in [1;W] \tag{5}$$

where $m(LV_{qz}, x_q; \beta_w^m)$ is the deterministic component of the measurement equation for statement $w$, $W$ is the number of the statements, $\upsilon_{qw}$ is the stochastic component of the measurement equation that is normally distributed with zero mean and variance $\sigma_w^2$.

Given the Likert scale structure we consider $I_{qw}$ as a discrete ordered value. Therefore, for five-point scale we have:

$$I = \begin{cases} j_1 & \text{if } I_{qw} < \tau_1 \\ j_2 & \text{if } \tau_1 \leq I_{qw} < \tau_2 \\ j_3 & \text{if } \tau_2 \leq I_{qw} < \tau_3 \\ j_4 & \text{if } \tau_3 \leq I_{qw} < \tau_4 \\ j_5 & \text{if } I_{qw} \geq \tau_4 \end{cases} \tag{6}$$

where $\tau$ is the vector of parameters considering the symmetry of the distribution, i.e., $\tau_1 = -\Delta_1 - \Delta_2$, $\tau_2 = -\Delta_1$, $\tau_3 = \Delta_1$, $\tau_4 = \Delta_1 + \Delta_2$; $\Delta_1, \Delta_2$ are parameters to be estimated.

So, incorporating the latent variables based on the measurement equations into the conditional probability of choosing *CS* supply channel we have the following:

$$P_q(s|x_{st}, x_q, LV_{qz}, I_{qw}, \beta) = \int_{\alpha_q} \int_{LV} \prod_{t=1}^{T} \begin{pmatrix} p_{qt}(s|x_{st}, x_q, LV_{qz}, \alpha_q, \beta) \cdot e(I_{wq}|LV_{qz}, \beta_w^m, \sigma_w) \\ \cdot g(LV_{qz}|\beta_q^s, \rho_z, \sigma_s) \cdot \varphi(\alpha_q|R) \cdot dLV_{qz} \cdot d\alpha_q \end{pmatrix} \tag{7}$$

where $\rho_z$ is the factor loading, $e(I_{wq}|LV_{qz}, \beta_w^m, \sigma_{I_w})$ is the density of distribution $I_{wq}$, $g(LV_{qz}|\beta_q^s, \rho_z, \sigma_s)$ is the density of distribution $LV_{qz}$, $x_q$ is the socio-economic attribute for consumer $q$, $x_{st}$ is the supply channel $s$ specific attribute within the choice task $t$, $T$ is the number of the choice tasks presented to every respondent within the stated-preference survey, $p_{qt}$ is the kernel of the conditional probability to be estimated based on the mixture of the logit given the agent effect:

$$p_q\left(s \mid x_{st}, LV_{qz}, \alpha_q, \beta\right) = \int_{\alpha_q} \prod_{t=1}^{T} \frac{\exp(U_{qst})}{\sum_{s \in S} \exp(U_{qst})} \cdot \varphi(\alpha_q \mid R) \cdot d\alpha_q \quad (8)$$

$$\alpha_q \sim N(0, \Sigma) \quad (9)$$

where $S$ is the number of the considered supply channels for e-groceries; $R$ is the number of draws.

The utilities $U_s$ for the alternative supply channels are formed as follows:

$$U_{qCS} = ASC_{CS} + X_s \cdot \beta_s + LV_q \cdot \beta_{LV} + \varepsilon_{qCS} \quad (10)$$

$$U_{qCC} = X_s \cdot \beta_s + \varepsilon_{qCC} \quad (11)$$

$$U_{qPS} = ASC_{PS} + X_q^* \cdot \beta_q^* + \varepsilon_{qPS} \quad (12)$$

where $U_{qPS}$ is the perceived utility by $q$ consumer when CS channel is chosen, $U_{qCC}$ is the perceived utility by $q$ consumer when CC channel is chosen, $U_{qPS}$ is the perceived utility by $q$ consumer when physical store for groceries is chosen representing the rejection of online services, $ASC_{CS}$ is the alternative specific constant for CS alternative, $ASC_{PS}$ is the alternative specific constant for physical store alternative, $X_s$ is the vector of attributes describing specific features of the supply channel $s$, $\beta_s$ is the vector of the parameters for the channel specific attributes, $\beta_{LV}$ is the vector of unknown parameters related to the latent variables of the utility, $X_q^*$ is the number of the children in the household for respondent $q$, $\beta_q^*$ is the number of children parameter.

*3.3 Stated-preference experiment design*

Unlike the revealed-preference survey, the stated-preference experiment provides the possibility to go beyond the current choice situation (Ortuzar and Willumsen, 2011). Within the case of new supply option deployment, the stated-preference survey technique becomes crucial to reveal possible choices and behaviors of the consumers. The methodology of the stated-preference experiments is based on the experimental design methodology (Fisher, 1935) and well adapted to the marketing and transportation fields (Louviere and Hensher, 1983; Kroes and Sheldon, 1988).

Within the first step to developing the choice sets, the attributes and their levels of variation in the full factorial design should be defined (Louviere et al., 2000). Given formalized trade-off conditions in section 3.1, the considered attributes and their levels are presented in Table 2. The levels for "delivery cost" attribute for CS option are calculated with 20% extra based on existing rates for e-groceries delivery by the commercial carriers in Ukraine (to be considered as a case study).

Hence, according to attribute levels we can estimate the full experimental plan that should be orthogonal to reduce the intercorrelation between attributes and provide appropriate statistical properties of the results (Louviere et al., 2000). The dimension of the full factorial experimental plan within studied case is evaluated as:

$$n = \prod_{m \in M} m^{k_m} \quad (13)$$

where $n$ is the number of the choice sets in the full experimental plan, $m$ is the considered option of the attribute's levels (two, three or more), $k_m$ is the number of the attributes for variation level $m$, $M$ is the final number of the attributes' variation levels within the experimental plan.

Table 2. Attributes and their levels

| Attribute | CS-based supply channel | CC-based supply channel |
|---|---|---|
| Delivery cost | 1st level – 60 UAH* (2.14 USD) | 1st level – 50 UAH (1.79 USD) |
| | 2nd level – 90 UAH (3.21 USD) | 2nd level – 75 UAH (2.68 USD) |
| | 3rd level – 120 UAH (4.29 USD) | 3rd level – 100 UAH (3.57 USD) |
| Delivery time | 1st level – less than 3 hours | 1st level – less than 6 hours |
| | 2nd level – 3-6 hours | 2nd level – 6-12 hours |
| | 3rd level – 6-9 hours | 3rd level – 12-24 hours |
| | 4th level – more than 9 hours | 4th level – more than 24 hours |
| Ecological contribution ($CO_2$ reduction) | Binary | - |
| Flexibility of the delivery | Binary | - |

*Note: Currency exchange course is 27.9 UAH to 1 USD, winter 2021

Given defined levels of the attributes, the following dimension of the full factorial experimental plan is defined: $n = 2^2 \cdot 3^2 \cdot 4^2 = 576$ choice sets. Such size of choices can be presented to the respondents demanding the formation of quite a significant sample to cover every choice set by several answers to provide the variability needed. To reduce the sample size, the fractional design should be implemented (Hensher, 1994). To do that, the AlgDesign package (Wheeler, 2004) for R software with Federov's fractional design algorithm is used. It allows obtaining the fractional with maximized determinant, which in such a way gives the minimal compounding matrix. The blocking procedure is required as respondents should face only some choice tasks from the fractional experimental plan (Hensher, 1994), and every block should represent the fractional plan efficiently. The critical question, in this case, is the fraction's size and the number of the choice sets within one block.

To solve this issue, Orme's (1998) methodology is used, allowing evaluation of the sample size needed from the prospect of choice set dimension and planned tasks per respondent. As the research team implements the data collection without funding, every respondent is expected to face a high number of choice tasks, namely nine games within the planned fraction size of 54 choice sets. Based on Orme (1998), such conditions guarantee the sample needed of 200-250 respondents, which matches the survey volunteering participation conditions. Given nine tasks per respondent, such design provides a panel effect that is planned to be accounted for during the discrete choice modeling. In this case, the research team traded off their efforts for data collection, i.e., covering as many as possible the number of games with a restricted number of respondents and serial correlation problems within the collected dataset (Wooldridge, 2005; Danalet et al., 2016).

## 4. Results

*4.1 Survey and sample description*

The survey was implemented in Kharkiv-City (Ukraine) which is considered as a case study. The online channel was used to collect the behavioral data with voluntarily conditions for the participants. The questionnaire was developed in Google Form in Ukrainian language and administered by research team. Given 54 choice sets presented by 6 blocks, the final number of Google Forms is six. The questionnaire contains four sections, where the socio-economic data on the respondents is collected within the *first section*. Within the *second section* 15 statements were presented to the respondents for their evaluation using five-point Likert scale, where five points reflect a strong agree/impact, and one point reflects no impact/disagree. These statements indirectly estimate people's attitudes towards online shopping and in-store shopping, their readiness to trust the groceries to occasional/not official carriers, flexibility features of the delivery and ecological concerns. The statements are the following (adapted from Kitamura et al. (1997) and Mokhtarian et al. (2009)):

- F1 – *The cost of online delivery is too high* ("–" online shopping);
- F2 – *The home delivery saves me time* ("+" online shopping);
- F3 – *Online shopping is too complicated* ("–" online shopping);
- F4 – *I like to visit stores and malls* ("–" online shopping);
- F5 – *The risk of purchasing a low-quality product is the main reason for shopping in a store* ("–" online shopping);
- F6 – *Delivery flexibility like delivery time, place etc. is important for me* ("+" crowd delivery);
- F7 – *I am concerned about the security and integrity of the delivery* ("–" crowd delivery);
- F8 – *I prefer social contacts when shopping in stores* ("–" online shopping);
- F9 – *Shopping in stores is very tiring* ("+" online shopping);
- F10 – *I trust only official representatives* ("–" crowd delivery);
- F11 – *The crowd-carrier is more reliable as he is personally responsible for the goods* ("+" crowd-shipping);
- F12 – *Crowd-shipping allows you to reduce emissions and air pollution* ("+" environmental impact);
- F13 – *I am not concerned about air pollution* ("–" environmental impact);
- F14 – *I travel green for the environment* ("+" environmental impact);
- F15 – *There are more important issues than environmental protection* ("–" environmental impact).

The *third section* of the questionnaire contains the choice set tasks that are introduced to the respondents, i.e., nine choice sets per participant. The delivery costs and times for the commercial carrier option were taken from the websites of the e-grocery providers in Kharkiv as "ROST," "Klass," "Spar," and "Metro." According to the stated-preference design presented in the previous section, the delivery costs and times for the CS deliveries were generated. An example of the choice set is depicted in Fig. 4.

| Attribute | | Online Commercial Carrier | Online Crowd-shipping Carrier | Physical store |
|---|---|---|---|---|
| Delivery cost | 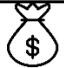 | **70 UAH** | **90 UAH** | |
| Delivery time | 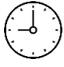 | **6 hours** or less | **3 hours** or less | |
| Ecological contribution | 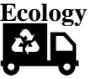 | **No reduction of $CO_2$** | **Reduction of $CO_2$** | |
| Flexible delivery | 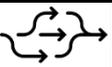 | **NO** | **YES** | |
| YOUR CHOICE | | ☐ | ☐ | ☐ |

Figure 4. An example of the choice set for groceries supplies

The *fourth section* comprises questions on the supply side of the CS, namely, what means of transport people prefer to use if they would act like a crowd-shipping courier and what remuneration value they would like to be paid for CS service. Given that, the behavioral data for demand and supply CS subsystems can be collected to reveal the delivery cost issue from the prospect of the client and service provider. The findings obtained by Gatta et al. (2019) for Italy and Wicaksono et al. (2021) for the Netherlands can be compared with this study's findings. Given the difference in the economies

12and service types, it would be interesting to reveal how people perceive the value of their time from the prospect of the demand and supply sides.

To disseminate the questionnaire, the master's students from the "Smart Transport and Logistics for Cities – SmaLog" programme (funded by Erasmus +) at O.M. Beketov National University of Urban Economy in Kharkiv (Ukraine), 2021 graduation year were involved. The research team aimed to reach a balance in the number of respondents faced with every block within the experiment. To have that the Google Forms were rotated according to their order from one to six for every new respondent. In such a way the similarity with the uniform distribution of the blocks number within the sample was reached. The snowball technique (Biernacki and Waldorf, 1981) was used for sampling to be effective under a low budget constraint. As a result, the survey covered 287 persons, and after data cleaning, the final sample was 249 respondents. The breakdown of the sample is presented in Table 3.

Table 3. The breakdown of the sample

| Attribute | Frequency | Sample, % | Population, % |
|---|---|---|---|
| Age | | | |
|   18 – 24 | 81 | 32.53% | 14.00% |
|   25 – 34 | 58 | 23.30% | 18.00% |
|   35 – 44 | 62 | 24.90% | 19.00% |
|   $\geq 45$ | 48 | 19.28% | 49.00% |
| Gender | | | |
|   Female | 121 | 48.59% | 53.64% |
|   Male | 128 | 51.41% | 46.36% |
| Household size | | | |
|   1 person | 55 | 22.09% | 24.90% |
|   2 persons | 50 | 20.08% | 33.60% |
|   3 persons | 78 | 31.33% | 26.80% |
|   4 persons and more | 48 | 26.51% | 14.70% |
| Car availability in household | | | |
|   Yes | 166 | 66.67% | NA* |
|   No | 83 | 33.33% | |
| Monthly personal wage, UAH** | | | |
|   < 5,000 | 28 | 11.24% | 11.80% |
|   5,000 – 9,999 | 83 | 33.33% | 19.70% |
|   10,000 – 19,999 | 80 | 32.13% | 35.90% |
|   20,000 – 29,999 | 30 | 12.05% | 23.10% |
|   30,000 – 39,999 | 6 | 2.41% | 5.30% |
|   40,000 – 49,999 | 9 | 3.62% | 2.60% |
|   $\geq 50,000$ | 13 | 5.22% | 1.60% |
| Employment status | | | |
|   Full-time | 155 | 62.25% | |
|   Part-time | 52 | 20.88% | |
|   Unemployed | 3 | 1.21% | NA |
|   Housekeeper | 15 | 6.02% | |
|   Student | 24 | 9.64% | |

Note: *NA – Not Available
    **Currency exchange course is 27.9 UAH to 1 USD, winter 2021

The sample comparison with the population shows their close similarity in terms of personal wage, household size, and gender. We did not reach a similarity in age attribute, namely for the category "more than 45 years." The focus on online services users can be explained by an effect as the younger generations are considered as early adapters for new services (Malokin et al., 2021). The being deployed SC-based service for e-groceries are considered such a case.

*4.2 Descriptive statistics*

Given the considered trade-off conditions, the descriptive statistics on cost and time attributes should be analyzed first. Most of the results depicted in this subchapter were collected within the fourth section of the questionnaire where people were asked to evaluate the cost and time attributes from the prospect of the consumer (demand-side) and service provider (supply-side). In the beginning of fourth section every respondent were asked to evaluate using importance of the delivery service attributes using 4-point scale, i.e., "Not important" – 1 point, "Very important" – 4 points. The results of the evaluation are depicted in Fig. 5. The *ecological contribution* has been revealed as less important among considered attributes which is expected as consumers first of all trade off cost and time while choosing the shopping and delivery channels (Rossolov, 2021; Meister et al., 2023). In addition, the revealed importance of *delivery flexibility* with almost equal estimations for *delivery time* and *delivery cost* highlights the consumers' behavior on willingness to manage and adjust their delivery process while it is being implemented.

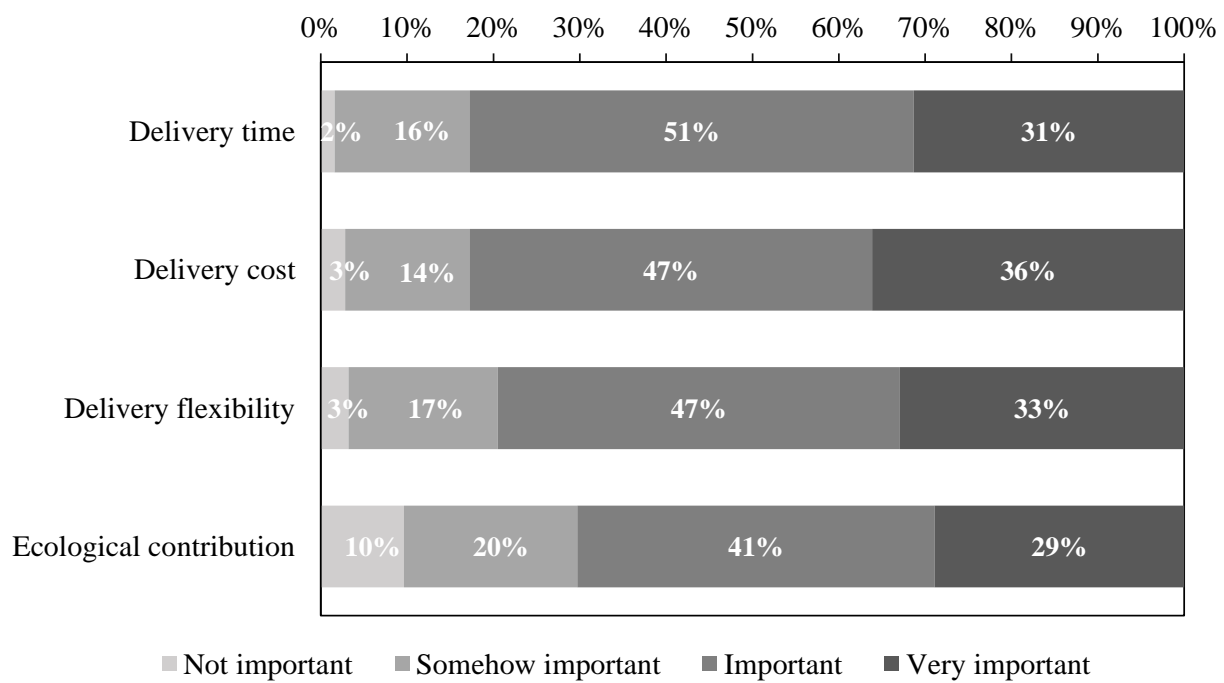

Figure 5. Ordered estimations on attributes' importance for e-groceries deliveries

Along with that, the respondents were asked to pick the remuneration for CS in range 50 – 120 UAH considering two points of view: consumer-side and courier-side. As a result, we detected a contradicting behavior when comparing the delivery remuneration for the demand and supply sides (Fig. 6). Consumers aged 18-24 and 25-34 are ready to pay less for CS than they expect to earn for it if they implement the delivery. The supposed remuneration (average estimates) is the highest for people aged 25-34 and equals to 85.43 UAH[1]. Meanwhile, with an increase in the age category, i.e., 35 years and more, a reduction in the *expected* remuneration for CS delivery (supply-side) is observed, namely, 81.45 UAH and 78.96 UAH, accordingly, for the 35-44 and more than 45 years old groups. What is more important is that people aged more than 35 are willing to pay more for CS service (demand side) then they expect for the remuneration if they act as the couriers.

---

[1] Average monthly wage per person was 14,014.00 UAH in 2021, Ukrainian State Statistic Services Service's report 2022
https://ukrstat.gov.ua/operativ/menu/infografika/2022/o_soc_ek_Ukr/01_2022_u.pdf

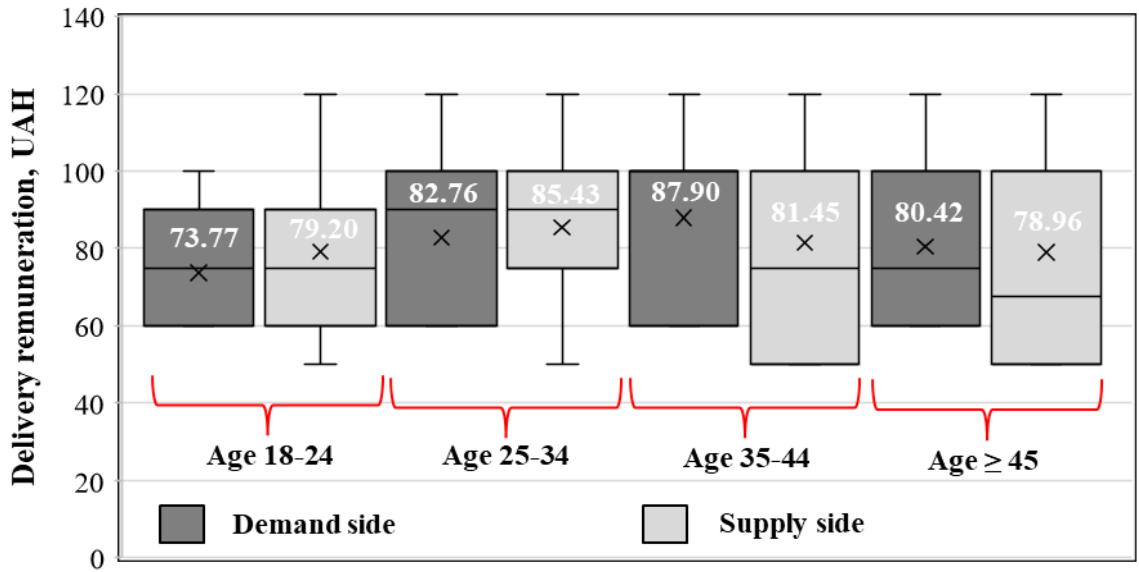

Figure 6. Acceptable remuneration for CS deliveries from demand and supply sides perspective

Hence, two types of behavior can be distinguished. The first one is oriented on *earning more* as possible money for the delivery and pay less when people act as the users of the CS services. Such behavior is relevant to the consumers aged up to 35 years old. The second behavior supposes the less *financial orientation* for the CS if people function as the occasional couriers. People aged more than 35 years behave in such a way. We assume they are less sensitive to the money earned based on CS due to income stability.

The next aspect which is of great interest is the mode preferred for the CS delivery. Given considered Kharkiv City, where the public transport is diverse in terms of spatial and modes available, i.e., subway, tram, trolleybus, bus, the occasional currier have a list of alternatives for implementing sustainable CS delivery. On the other hand, the level of motorization in the last years tended to grow and reached 550 vehicles per 1000 inhabitants in 2021, indicating significant attraction to car-based mobility. In such conditions, the willingness to use specific modes for CS deliveries concerning car ownership was evaluated. The obtained results are presented in Fig. 7.

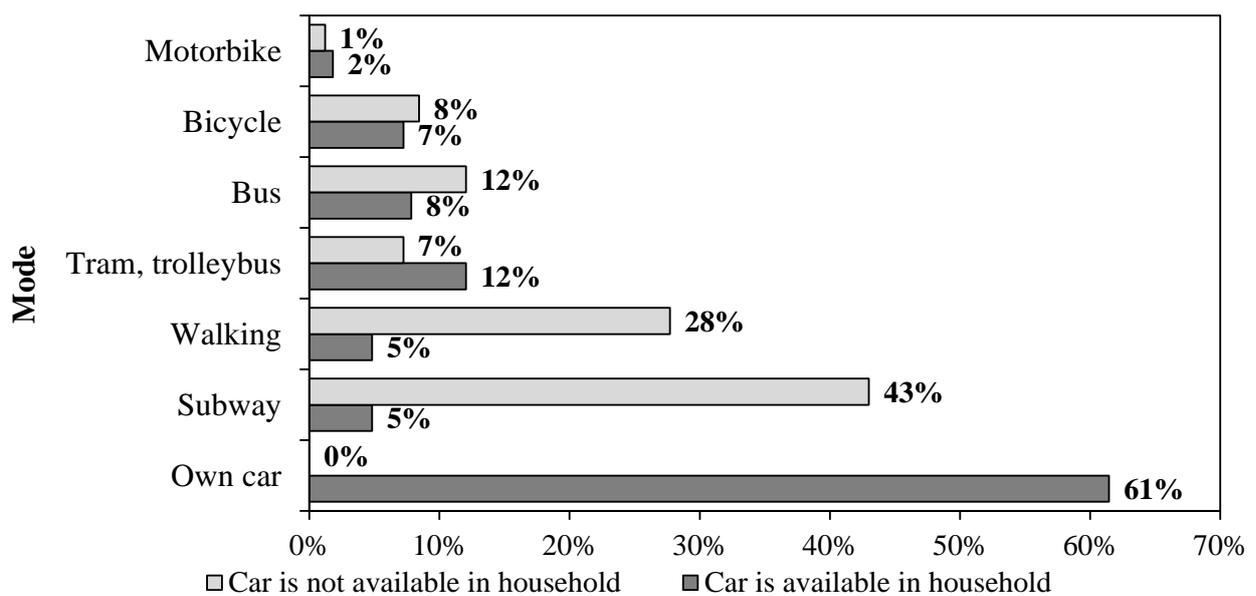

Figure 7. Modal split for the CS deliveries (readiness to provide opinion)

As a result, for non-car commuters, the subway and walking are more preferable modes for CS, respectively, with 43% and 28% of respondents not possessing a car. Additionally, 12% of respondents are ready to use a bus, and 8% – a bicycle. The green ground transit modes as tram and trolleybus, attracted only 7% of non-car commuters. On the other hand, car ownership may dramatically influence on mode choice for CS deliveries. For instance, 61% of respondents that possess a car in their household would prefer implementing CS delivery by car, which contradicts CS's main idea – reduction a carbon footprint. All other modes would be used barely and might be presented by shares from 5% to 12%.

With respect to the mode choice by the potential CS couriers, the detour becomes crucial as it reveals the possible impact on traffic if the car is used for the delivery and provides the information on spatial coverage of CS deliveries. Given stated mode for CS delivery, the respondents were asked to indicate the acceptable maximum detour time in range from 15 to 60 minutes. The distribution of the revealed detour times was summarized in box and whisker plot (Fig. 8).

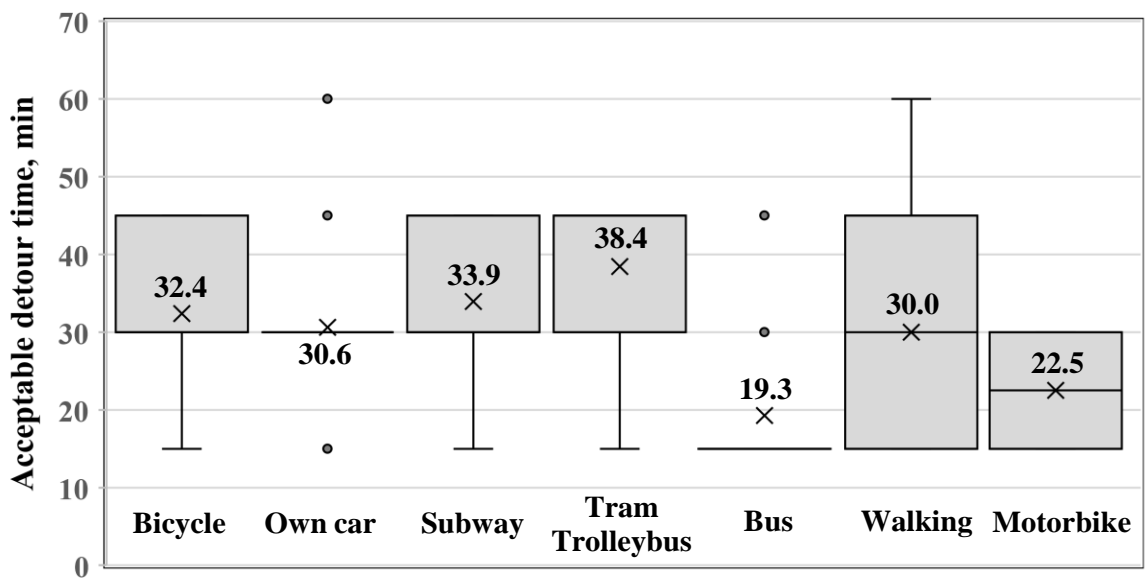

Figure 8. Maximum acceptable detour time for providing CS service

The maximum acceptable detour time for CS deliveries was estimated for tram and trolleybus with a mean of 38.4 minutes, and the minimum – for bus with a mean of 19.3 minutes. But given the small shares of these modes, it is more interesting to emphasize the results for the most attractive modes, i.e., car and subway. The average values of the acceptable detour times for them are almost equal, 30.6 minutes for a car and 33.9 minutes for a subway. But for the car, most of the estimates are firmly located near the mean value, which resulted in the absence of the quartile fields. Such estimations can be explained by the perceived high value of time (Antoniou et al., 2007; Schmid et al., 2019) for car commuters which strongly determines their willingness to make the detour during their trip. Interestingly, the acceptable detour time for the subway is not much bigger than for the car, with a mean value of 33.9 minutes. It can be assumed the value of time for occasional couriers, that would use the subway for delivery, would be higher than for ordinal subway commuters. Such an assumption is made based on the fact that couriers should bear the additional package of food to be delivered to the consumer. In this case, the perceived value of time during the delivery by subway may increase. It should be mentioned that this is only an assumption that will be studied in depth in future.

The revealed CS-based delivery patterns are only the first steps made in direction of mode choice behavior for CS deliveries distinguishing the future research needed. For instance, it is worth to be mentioned that revealed detour times give the information on possible negative consequences due to the CS deliveries if private car is used. Given the average speed of 20 km/h in peak hours the car-based CS delivery could produce an additional 10.2 km, according to the mean of the detour time revealed. This is just a hypothetical estimate, but even with such an analytical analysis, it can be assumed that the negative consequences of CS-based deliveries are real.

*4.4 Exploratory factor analysis*

The factors extraction is implemented using factor analysis. Given used five-point Likert scale, the correlation between latent factors and the score made by the respondents can be estimated, resulting in factors extraction with relevant values of the loadings. Considering the methodology of the factor analysis (Harman, 1976), the eigenvalues were evaluated and using Kaiser criterion (Kaiser, 1958) the factors with eigenvalue less than 1.0 were excluded from the list. As a result, four factors were extracted. In the next step the varimax rotation was made to estimate the loadings for the factors. The loadings with value less than |0.4| were excluded from the list. After implementing this procedure, the results of the factor analysis obtained and summarized in Table 4. The described operations were made with STATISTICA 7.0 software.

The extracted factors describe the following behavior:
- Factor 1 is determined with seven statements that reflect the desire of end consumers to save time thanks to online orders, delivery with a flexible schedule, high reliability, and environmentally friendly;
- Factor 2 is characterized by statements on environmental indifference and fixation of end users on their own needs;
- Factor 3 reflects the orientation of end consumers towards purchasing food products in physical stores, which is due to the satisfaction of visiting them and inability to check the quality of the groceries before purchasing;
- Factor 4 characterizes the advantage of social contacts when visiting physical stores compared to the difficulty of using online services.

Table 4. Rotated factor loadings

| Statement's code | Extracted factors | | | |
| --- | --- | --- | --- | --- |
| | Factor 1 "Pro-Crowd" | Factor 2 "Pro-Official Carriers" | Factor 3 "Pro-Store" | Factor 4 "Social contacts" |
| F1 | -* | - | 0.58 | - |
| F2 | 0.66 | - | - | - |
| F3 | - | - | - | 0.80 |
| F4 | - | - | 0.59 | 0.55 |
| F5 | - | - | 0.70 | - |
| F6 | 0.71 | - | - | - |
| F7 | 0.68 | - | - | - |
| F8 | - | - | - | 0.85 |
| F9 | 0.56 | - | -0.40 | - |
| F10 | - | 0.67 | - | - |
| F11 | 0.80 | - | - | - |
| F12 | 0.74 | - | - | - |
| F13 | - | 0.90 | - | - |
| F14 | 0.60 | - | - | 0.46 |
| F15 | - | 0.84 | - | - |

Note: * Factor loading is less than |0.4|

The extracted latent factor #1 was incorporated into the discrete choice model based on F2, F6, F7, F9, F11, F12 and F14. When the parameters for the measurement equations will be estimated one of the statements (factors) should be normalized to zero. Given that, this procedure was applied to F2.

4.5 Discrete choice modeling results

The hybrid choice model was estimated in Biogeme (Bierlaire, 2020). The simultaneous hybrid model estimation was implemented to incorporate in one likelihood function direct discrete choice model (so-called kernel) and latent variable based on structural and measurement equations. To account for the serial correlation effect due to the panel data along with a necessity to randomize the latent variable, the Monte Carlo draws technique is used. Having that, it is possible to estimate the integral of the conditional probability by simulating the likelihood function (Revelt and Train, 1998; Fosgerau and Bierlaire, 2007). The Normal distribution is chosen as the type of the probability distribution function. The structural equation is formed based on the factor analysis results for "Pro-Crowd" factor. During the model estimation some of the parameters were scaled, i.e., delivery time scaled to 10, delivery cost – to 100, income – to 10,000. Moreover, to capture the heterogeneity the ecological and delivery flexibility attributes were multiplied by income, resulted in the following expressions: $\beta_{CO_2} \cdot CO_2 \cdot Income\_scaled$, $\beta_{FLEX} \cdot Flex \cdot Income\_scaled$. The results of the estimates evaluated for hybrid choice models are summarized in Table 5.

The signs of the choice model parameters are consistent with the trade-off conditions, except for the environmental impact, which is negative, contradicting the expected positive value. Since it is not statistically significant, the sign could invert due to a lack of necessary observations. Also, such a behavior can be explained by the fact that most of the consumers with cars are ready to use this mode in case they would act as CS's occasional carrier. Given that, we have the contradicting effect, which can explain why the ecological aspect becomes less important for online grocery shoppers in some situations.

Based on the obtained estimates on the first step the willingness-to-pay (WTP) is estimated to reveal the willingness of the end-consumer to pay an additional cost for some modification in the service (Train, 2003). In the considered case, the focus is given to the trade-off between delivery time and delivery cost to account for service quality improvement by reducing delivery time. Hence, the WTP for consumer $q$ can be defined with respect to the scaling procedure for the time and cost attributes as follows:

$$WTP = \frac{\partial V_{qs}/\partial TIME_{qs}}{\partial V_{qs}/\partial COST_{qs}} = \frac{\beta_{TIME(qs)}}{\beta_{COST(qs)}} \cdot 10, \left[\frac{UAH}{hour}\right] \tag{14}$$

Given the sample, the estimation of WTP is being implemented based on the estimates for $q$ respondents surveyed. In this case, *eq*. (14) transforms into $\frac{\partial V_s/\partial TIME_s}{\partial V_s/\partial COST_s} = \frac{\beta_{TIME(s)}}{\beta_{COST(s)}} \cdot 10$ and the mean values of WTP are presented in Table 6.

Table 6. The values of WTP for alternative delivery channels

| Groceries delivery channel | WTP, UAH/hour |
|---|---|
| Crowd-shipping | 7.47 |
| Commercial carrier | 2.01 |

The obtained estimates for WTP reproduce the value of delivery time savings which is significantly higher for CS delivery channel. It can be characterized by the fact that people which have chosen CS expect more quality from the service and ready to make an extra payment for that. The flexibility of the delivery revealed as the significant attribute that contributes to the perceived quality of the e-groceries delivery too. Given the fact that the data collection was made during the COVID-19 pandemic, when the number of infection cases in Ukraine was ~ 10K per day (WHO, 2021), the revealed findings show that even in crisis time, the quality of the delivery service remained to be important for the online grocery shoppers and they are ready to pay extra money for that.

Table 5. Hybrid choice model estimation results

| Attribute's code | Value | Robust standard error | Robust t-test |
|---|---|---|---|
| *Choice model* | | | |
| $ASC_{CROWD}$ | -0.112 | 0.34 | -0.33 |
| $ASC_{STORE}$ | -1.91*** | 0.256 | -7.47 |
| $\beta_{CHILDREN}$ | -0.288*** | 0.0625 | -4.61 |
| $\beta_{CO2}$ | -0.00477 | 0.0427 | -0.112 |
| $\beta_{COST\_COMER}$ | -3.05*** | 0.308 | -9.9 |
| $\beta_{COST\_CROWD}$ | -1.95*** | 0.201 | -9.72 |
| $\beta_{FLEX}$ | 0.125*** | 0.0428 | 2.93 |
| $\beta_{TIME\_COMER}$ | -0.621*** | 0.084 | -7.4 |
| $\beta_{TIME\_CROWD}$ | -1.44*** | 0.207 | -6.96 |
| $\beta_{PRO\text{-}CROWD}$ | 0.138* | 0.056 | 2.47 |
| $\sigma_\alpha$ | 0.528*** | 0.058 | 9.09 |
| *Measurement equations* | | | |
| $\beta_{0(F6)}$ | -0.495*** | 0.0958 | -5.17 |
| $\beta_{0(F7)}$ | -0.0526 | 0.068 | -0.773 |
| $\beta_{0(F9)}$ | -1.4*** | 0.0924 | -15.1 |
| $\beta_{0(F11)}$ | -0.915*** | 0.0872 | -10.5 |
| $\beta_{0(F12)}$ | -1.12*** | 0.104 | -10.7 |
| $\beta_{0(F14)}$ | -0.936*** | 0.1 | -9.33 |
| $\beta_{F6}$ | 1.07*** | 0.0582 | 18.3 |
| $\beta_{F7}$ | 0.804*** | 0.0406 | 19.8 |
| $\beta_{F9}$ | 0.952*** | 0.0538 | 17.7 |
| $\beta_{F11}$ | 1.19*** | 0.053 | 22.5 |
| $\beta_{F12}$ | 1.33*** | 0.0646 | 20.7 |
| $\beta_{F14}$ | 1.05*** | 0.0589 | 17.9 |
| $\sigma^*_{F6}$ | 0.961*** | 0.0485 | 19.8 |
| $\sigma^*_{F7}$ | 0.697*** | 0.0328 | 21.3 |
| $\sigma^*_{F9}$ | 1.29*** | 0.0647 | 20 |
| $\sigma^*_{F11}$ | 0.626*** | 0.0391 | 16 |
| $\sigma^*_{F12}$ | 0.773*** | 0.0479 | 16.1 |
| $\sigma^*_{F14}$ | 0.922*** | 0.0492 | 18.8 |
| $\Delta_1$ | 0.653*** | 0.0324 | 20.2 |
| $\Delta_2$ | 0.752*** | 0.0336 | 22.4 |
| *Structural equation* | | | |
| $\beta_{0(structural\ equation)}$ | 2.15*** | 0.108 | 19.9 |
| $\beta_{INCOME}$ | -0.182*** | 0.0237 | -7.67 |
| $\beta_{AGE\_MORE\_30}$ | 0.424*** | 0.0659 | 6.43 |
| $B_{PART.TIME.OCCUPATION}$ | 0.17*** | 0.0534 | 3.18 |
| $B_{HIGH.EDUCATION}$ | -0.55*** | 0.0578 | -9.52 |
| $B_{MORE.MEMBERS}$ | 0.222*** | 0.0495 | 4.49 |
| $\beta_{MALE}$ | -0.147*** | 0.0523 | -2.82 |
| $\beta_{NoCAR}$ | 0.0808 | 0.05 | 1.62 |
| $\sigma_S$ | 0.815*** | 0.064 | 12.7 |
| *Model's general information* | | | |
| Number of draws | | 1000 | |
| $LL_{null}$ | | -30378.59 | |
| $LL_{final}$ | | -18629.71 | |

Note: Robust p-values ***: $p < 0.005$, **: $p < 0.01$, *: $p < 0.05$.

## 5. Discussion and study's limitations

Having the descriptive, factor, and discrete choice analyses results, the following findings for the CS-based e-grocery deliveries in terms of the consumers' behavior can be depicted:

(i) The revealed latent variable "Pro-Crowd" contributes positively to the utility of the CS delivery channel representing the socio-demographic groups of e-shoppers with positive pro-crowd-shipping attitudes. Females are more pro-CS oriented than males, which is complemented by the part-time occupation status, which is more common for women. Consumers over 30 years old have a pro-CS delivery service behavior characterized by their welfare stability and readiness to pay extra money for the higher quality of the delivery. Besides the above depicted, the family size matters, providing more pro-CS-oriented behavior when the number of household members increases.

(ii) The factor analysis results indicate the necessity to make the last-mile e-grocery logistics more flexible and eco-friendlier. The possibility to adjust the time and place to pick up the package with the groceries is a key option distinguishing e-shoppers between crowd-shipping and commercial carrier oriented. What is more important and should be emphasized is the personal responsibility of the CS courier for the delivery compared with CC. This factor was revealed with the highest loading value, meaning that this delivery service feature is on top of the factors list for the consumers. Given that, it can be stated that the reliability of the e-groceries delivery complements the WTP's findings showing the interconnection between the perceived value of the delivery time and the security of the delivery services.

(iii) The estimated WTP for CS is 270% higher than for the CC service channel, which is far larger than was expected. Such phenomenon is partly explained above by the exploratory factor analysis results. In this case, the sensitivity to the delivery time change is higher for CS services, and the delivery conditions are more important than for CC. Given that, the mode used for the CS delivery can significantly affect the attractiveness of the CS services. For instance, using the private car, a desirable option for car owners (Fig. 7), might not only negatively contribute to traffic and environmental pollution but also could be unreliable in terms of delivery times with a high likelihood of delivery delays.

(iv) The revealed average values for remuneration that can be paid by the consumers and expected to be paid for them if they act as the occasional courier highlight the threshold in the age attribute. Having the sample's strata on age, people from 18 to 34 years old behave in a more money-earning shape as they expect to have a higher remuneration for CS delivery service that they are ready to pay for it. In turn, people from the 35 to more than 45 years group are more likely to provide the CS service with a lower remuneration rate than they are ready to pay for it. Such behavior is more pronounced for the age group "35-44" years old as they have the highest spread between the remuneration rates of 6.45 UAH (based on the average values of the estimates). This behavior is regarded as more relevant to the crowd-sourcing paradigm based on the people's willingness to contribute to society first, and only after that the money stimulus comes.

The key findings of this paper can be compared with previous studies. Having various approaches to data collection, i.e., revealed and stated-preference methods, case studies for different economies and CS service levels of development, the commonalities and differences can be revealed. Table 7 contains the summary of this study and the earlier researchers made within behavioral aspects of CS-based services. The attributes' considered were distinguished into two groups: end-consumers characteristics and alternative delivery channel parameters. The reference of some continuous attributes as income, delivery time and cost were done according to their impact on the perceived utility for the choice made. Meaning for instance, if an increase in the delivery cost will reduce the readiness of consumers to choose CS-based services, this attribute should be referred to as a "negative" impact – the same stands for other continuous attributes.

Table 7. Comparison of statistically significant attributes of CS-based deliveries choices (demand-side)

| Attributes' type | Developed economies (literature) | | Developing economies (this study) | |
| --- | --- | --- | --- | --- |
| | Positive | Negative | Positive | Negative |
| End-consumers socio-demographic characteristics | <ul><li>income</li><li>full-time employment</li></ul> | <ul><li>income greater than 59 000 USD</li><li>low education level</li><li>part-time employment</li><li>senior age of consumers</li></ul> | <ul><li>part-time employment</li><li>age more than 30 years</li><li>family size</li></ul> | <ul><li>income</li><li>high education</li><li>male</li></ul> |
| Channels attributes | <ul><li>flexible delivery</li><li>eco-friendly service/$CO_2$ reduction</li><li>driver high rank</li><li>parcel tracking</li></ul> | <ul><li>delivery time</li><li>delivery cost</li><li>detour time</li></ul> | <ul><li>flexible delivery</li></ul> | <ul><li>delivery time</li><li>delivery cost</li></ul> |

Having that, the similarities between developed and developing economies were revealed for channel-specific attributes, i.e., delivery cost, time, and flexibility options. These are expected results as these attributes reflect the trade-off process for quality (low delivery time) and cost (low delivery cost) of the service provided. Delivery flexibility also impacts the consumer's choice behavior for both developed and developing economies in the same way. The difference between consumer's behavior in developed and developing economies was estimated for specific socio-demographic attributes. Thus, personal income attributes for developed countries defined as pro-CS service parameter, meaning its increase will positively impact on readiness to use CS-based services (Devari et al., 2017; Punel and Stathopoulos, 2017). In contrast to that, consumer income increases in developing economies might lead to a reduction in the likelihood of CS-based groceries supplies being chosen (this study findings). Such an effect can be explained that consumers will tend to use the commercial carrier services. This finding correlates with Punel et al. (2018) study results as they estimated people with income of more than 59,000 USD to be non-CS-users. Other contradictions in the obtained results were detected for employment status and age of the consumers. Thus, according to Punel and Stathopoulos, 2017, and Punel et al. (2018), people with full-time occupations tend to use CS-based services. While for developing economies, part-time employees are more likely to be CS users. As for age, Gatta et al. (2019) defined that aged people are not ready to utilize CS-based services. This study revealed for developing economies that people up to 30 years old are not considered themselves as potential CS users.

Along with depicted findings, the study has limitations which are caused by the sample size and its slight differences with the total population characteristics. Given that, the revealed features of the consumers' behavior in terms of the delivery channel choice and perceived quality of the CS service can be considered from the case study point of view and should be checked and confirmed based on the more wide-scale survey. But we would like to assure the reader of the possibility of using the proposed methodology to reveal the readiness of e-shoppers to pay extra money for the higher quality of the e-grocery delivery based on CS service.

## 6. Concluding remarks

The crowd-shipping is one of the possible ways to reduce the negative externalities of last-mile logistics that become crucial in the light of dynamic e-commerce development. The intensively growing e-groceries are one of the main e-commerce areas where the crowd-shipping paradigm is expected to be efficiently implemented. Given this process is relevant for both developed and developing economies because of the COVID-19 factor, the behavioral study is the basic step that should be done to provide attractive and consumer-oriented CS delivery service for e-shoppers. This paper considers the delivery channel choice problem with crowd-shipping and commercial carriers as the main alternatives for e-groceries. The study is based on the assumption that promotion of the CS-based delivery service can be implemented through the increase in the delivery rates to engage the occasional people in the delivery process along with the provision of flexible and instant delivery service to attract the e-shoppers.

To consider the complexity of behavior in the choice situations, the choice model has been enhanced with the latent variable describing the pro-crowd-sourcing behavior of specific socio-demographic groups. The field of the study mostly covers the demand-side of the CS delivery problem with partly investigation of the preference in mode choice and acceptable detour times related to the supply-side of the system. To reveal the e-shoppers behavior in regard to their readiness to use CS-based e-groceries the stated-preference method was used to develop the choice games and propose them to the respondents through the online survey. The hybrid choice model was estimated based on the data collected in one of the largest Ukrainian cities – Kharkiv City.

The study's key findings lie in the revealed WTP values for CS and commercial carrier delivery channels. It was shown that people are ready to pay extra money for CS delivery under conditions when this service is provided with higher quality. The instant feature of the deliveries (Dablanc et al., 2017), accompanied by the flexibility of the service in terms of the pick-up location and time adjustments, attracts e-shoppers more than usual commercial carrier services. These findings have the similarities with revealed behavior for the developed economies which reflects alike consumption features of people for both economies. But unlike from people in developed economies, part-time employed consumers are more likely to use CS-based services in developing economies. Along with that, the positive environmental contribution of CS deliveries was revealed as not valuable for the respondents in the developing economies. This fact explains why people who own a private car consider this mode most relevant for implementing CS deliveries acting as occasional couriers. Such an aspect is of great interest as CS should reduce the negative impact of last-mile deliveries on the environment but not make it worse. Learning from these results, the necessity of deeper evaluation of the mode choice behavior for CS deliveries is crucial. The planned follow up study will cover this issue.

**CRediT authorship contribution statement**

**Oleksandr Rossolov:** Conceptualization, Methodology, Data curation, Software, Writing – original draft, Editing. **Yusak O. Susilo:** Formal analysis, Writing – review & editing, Funding acquisition.

**Acknowledgments**

The study became possible in the context of its implementation and presentation in its current form thanks to the support of the DAVeMoS research group and the Institute for Transport Studies at the University of Natural Resources and Life Sciences, Vienna, which hosted Ukrainian scientists.

**Declaration of Competing Interest**

The authors declare that they have no known competing financial interests or personal relationships that could have appeared to influence the work reported in this paper.